\documentstyle[12pt]{article}

\catcode`\@=11

\begin{document}
%\draft

%\def\baselinestretch{1.2}

\catcode`\@=11

\def\be{\begin{equation}}
\def\ee{\end{equation}}
\def \lab {\label}
\def \del {\partial}
\def \bd {\bar \partial }
\def \na {\nabla}

\def \ha{{\textstyle{1\over 2}}}
\def \na {\nabla }
\def \D {\Delta}
\def \a {\alpha}
\def \b {\beta}
\def \chi {\chi}
\def \s {\sigma}
\def \p {\phi}
\def \m {\mu}
\def \n {\nu}
\def \vp {\varphi }

\def \t {\theta}
\def \td {\tilde }

\def\ci{\cite}
\def \la {\label}
\def \sm {$\s$-model\ }
\def \foot {\footnote }
\def \P {\Phi}
\def \o {\omega}
\def \inv {^{-1}}
\def \ov {\over }
\def \four{{\textstyle{1\over 4}}}
\def \fourth{{{1\over 4}}}
%%%%%%%%%%%%%%%%%%%%%%%%%%%%%%%%%%%%%%%%%%%%%%%%%%%%%%%%%%%%%%%
%%%%%%%%%%%%%%%%%%%%%%%%%%%%%%%%%%%%%%%%%%%%
\def\np {  Nucl. Phys. }
\def \pl { Phys. Lett. }
\def \mpl { Mod. Phys. Lett. }
\def \prl { Phys. Rev. Lett. }
\def \pr  { Phys. Rev. }
\def \ap  { Ann. Phys. }
\def \cmp { Commun. Math. Phys. }
\def \ijmp { Int. J. Mod. Phys. }
\def \ijmp { Int. J. Mod. Phys. }
\def \jmp { J. Math. Phys. }
\def \cqg {Class. Quant. Grav. }
%%%%%%%%%%%%%%%%%%%%%%%%%%%%%%%%%%%%%%%%%%%%%%%%%%%
\begin{titlepage}
\begin{flushright}
IASSNS-HEP-95/79\\
Imperial/TP/95-96/4\\
hep-th/9510097\\
October 1995
\end{flushright}
\begin{center}
{\Large\bf General class of  BPS saturated}\\
\vskip 0.2cm
{\Large\bf  dyonic black holes   }\\
\vskip 0.2cm
{\Large\bf  as exact superstring solutions}
%solutions of superstring theory
%{\Large\bf  toroidally compactified  heterotic string} \\
\vskip 1cm
{\bf Mirjam Cveti\v c\footnote{On sabbatic leave from  the University of
Pennsylvania. e-mail: cvetic@sns.ias.edu}}
\\
\vskip 0.2cm
{\it School of Natural Sciences}\\
{\it Institute for Advanced
Study, Princeton, NJ 08540, USA }
\vskip 0.2cm
and\\
\vskip 0.2cm
{\bf Arkady  Tseytlin\footnote{On leave from Lebedev Institute,
Moscow. e-mail: tseytlin@ic.ac.uk }}\\
\vskip 0.2cm
{\it Theoretical Physics Group, Blackett Laboratory }\\
{\it Imperial College,   London SW7 2BZ, U.K.}
\end{center}

\begin{abstract}
{We show that a four-parameter generating solution
for a general class of four-dimensional,
spherically-symmetric, static, dyonic BPS saturated
solutions of leading-order effective equations of
toroidally compactified heterotic string  theory
is an exact string solution. The corresponding
ten-dimensional background  defines a conformal sigma-model
 which is a particular case of a `chiral null model'
with curved `transverse' part. The exact conformal invariance
is a consequence of the chiral null structure of the `electric'
part of the model and the $N=4$ world-sheet supersymmetry of
its transverse `magnetic' part. The sigma-model action has
a remarkable covariance under both target space $T$-duality and the
electro-magnetic $S$-duality transformations,
and  it illustrates the relationship between string-string duality
in six dimensions and $S$-duality in four dimensions.
In general, there exists a large class of exact six-dimensional
superstring solutions described by chiral null models
with four-dimensional transverse parts  represented by $N=4$
supersymmetric $\sigma$-models  with metrics conformal
to hyper-K\"ahler ones.}

\end{abstract}
\end{titlepage}

\newpage

1.{\it \ Introduction.}\    Recent recognition \cite{HTI,WITTENII} that
 Bogomol'nyi-Prasad-Sommerfield
(BPS) saturated   states  should play an important
 role in  the full
non-perturbative dynamics of string theory,
 has triggered a renewed interest in
%revisiting old and constructing new
 supersymmetric  solutions   of different
effective field theories
 in  various  dimensions
(for a review, see \cite{Townsend} and references therein).
The  study of such
configurations may, in particular,
shed light  on  duality symmetry between certain strongly coupled
and  weakly coupled string vacua.

These backgrounds   have  minimal energy
in their class, {i.e.},
they  saturate the  Bogomol'nyi bound on the energy,
and
are usually  obtained  as extrema  of the
leading-order  effective string-theory
(supergravity)  actions.
%to the lowest order in $\alpha'$
%expansion.
These solutions  admit Killing spinors
%satisfy the
%Killing spinor equations,
and thus preserve some of the
supersymmetries.
% (which are obtained by setting to zero the
%supersymmetric variations of fermionic fields equal )
% in these bosonic configurations.

To interpret  such a  configuration   as  a solution of {\it string theory}
(and not just of supergravity theory)
and thus to achieve  an adequate  understanding of its  properties
and  its  role in the context of string theory,
one should identify  the corresponding conformal  two-dimensional (2d) field
theory.
The first step  in this direction is
to find  a  conformal 2d \sm
with couplings $(G_{\m\n},B_{\m\n},\P, ...)$
which reduce to the  background  fields
in the leading order in $\a'$, {i.e.},   which  extend
the leading-order background to  an exact string solution.

There  are  large classes of   conformal $\s$-models,
or exact  string solutions,
which actually retain their leading-order form: they
 are not modified by $\a'$-corrections
provided one uses a special scheme
(for a review, see \cite{TS}  and
references therein).
Such are
`chiral null models'  \cite{TH},
which, in particular,
  describe (after the Kaluza-Klein truncation)
%(assuming a Kaluza-Klein interpretation)-you can change this back
 %all known
supersymmetric extreme {\it electric}
four-dimensional
black-hole-type solutions
(see, e.g., \cite{HRT,TH,BEHRNDT2,BEHRNDT,KALLIND,HOS}).
 The special chiral null structure of the corresponding  \sm  implies that
these solutions
are exact  not only in the
type II superstring or heterotic string theory, but also in the
bosonic string theory \cite{TH}.
On the other hand,
 it is the extended supersymmetry that is crucial for the exactness of the
  extreme    {\it magnetic} black hole solutions, i.e.,  they are exact
only as
% type II superstring  solutions or  symmetrically
  heterotic or type II superstring solutions \cite{chs,HMON,dukh,nels,kalor}.

The aim  of the present  paper is to  show that
 a
 %general
 class of  four-dimensional
spherically-symmetric static  {\it dyonic }
BPS saturated    solutions   of
leading-order effective equations of
toroidally compactified
 heterotic
 % (or type II)  super
 string  theory (for a review, see \cite{CveticYoum}
and references therein)
 is  actually   a class of
{\it exact  string solutions}
 by presenting the corresponding
conformal $\s$-model.
By  embedding these
 solutions into
 ten-dimensional
string theory
we shall  find that the  resulting  superstring action
is a particular case of the   generalized
chiral  null model \cite{TSH,TH,TS}
with {\it curved} four-dimensional `transverse'  part.
%We  shall   describe  how  its special cases reduce to
%the  pure electric or pure magnetic
%solutions  discussed previously.
  Although  the general background  with the  non-vanishing
electric and magnetic charges  corresponds to a  model with
only  $N=1$ space-time supersymmetry, so  that  the  resulting
 \sm possesses only
$N=2$ world-sheet supersymmetry,
the chiral null structure
of the  `light-cone'  (`electric') part of the  \sm
action, along   with the extended  $N=4$ world-sheet supersymmetry\foot{In
what follows $N=4$ world-sheet supersymmetry  would mean $(4,0)$
supersymmetry in the heterotic string case and $(4,1)$ supersymmetry in the
type II (or `embedded' heterotic)  case.
}
of the  four-dimensional `transverse'   (`magnetic')  part,
 will suffice  to argue
that the  full 2d theory is conformal to all orders.
We shall  thus reach a  remarkable conclusion  that  the
electric and magnetic charges associated with  two
different compact Kaluza-Klein   dimensions    can be superposed
at the level of conformal $\s$-model without  need to introduce
extra terms in the
 2d action.
This is a non-trivial result,   given that the `dyonic'  \sm
does not  factorise into the sum of `electric'
and `magnetic'  parts since
 both   depend on the same radial coordinate.

\vskip 0.2cm

2.{\it \ General Dyonic BPS Saturated Black
 Hole Solution.}\  The supersymmetric
 backgrounds  we are going to  consider are  solutions of the
four-dimensional $N=4$ supersymmetric effective action of the heterotic string
 compactified on a six-torus (see,  { e.g.}, \cite{SEN2} and references
 therein).
  At generic points of the moduli space of
toroidally compactified heterotic string the set of BPS saturated,
supersymmetric,  static, spherically
symmetric configurations    is both $O(6,22,Z)$ and
$SL(2,Z)$  duality invariant.
 In  \cite{CVETICYOUM}  the explicit
form of a general class of such configuration with 56 charges subject
to one constraint was given.
 It corresponds to the states which can be obtained
by acting with  $SL(2,Z)$ transformations on configurations
 with {\it  zero axion  and
most general allowed dyonic charges}.

The latter set of
solutions  are  $O(6,22,Z)$ orbits of dyonic
configurations  whose `left'  as well as `right' electric
and magnetic charges are {\it orthogonal}, { i.e.},  light-like in the
$O(6,22,Z)$ sense.
It turns out \cite{CVETICYOUM} that the generating
solution for this set  is
 the one   which  has the  non-trivial
 scalar fields being   represented only  by the
diagonal  components of the internal metric  of the six-torus ($g_{mm}$,
$m=1,...,6$)
and the four-dimensional dilaton ($\p_4$),
whose asymptotic values may be  set equal  to one and zero, respectively.

 The explicit form of this  {\it generating solution}  depends
 on  four parameters:
two magnetic and   two electric charges.   The  magnetic  and electric
charges are associated with   two  {\it  different}  internal dimensions, i.e.,
two different  $U(1)$ groups.
The  magnetic (electric)
charges  correspond to
the Kaluza-Klein  vector  field $A^{(1) m}_{\vp}$
($A^{(1) n}_{t}$) and the  vector  field originating from the
antisymmetric tensor components
$A^{(2)}_{\vp m}$ ($A^{(2)}_{t n}$).
Here  the indices $t$ and $\vp$ refer to  the
time and polar angle  coordinates
of four-dimensional
 space-time and  $m\ne n$  are a pair of the  six-torus indices (for the
explicit
form of the
solution in terms of  the four-dimensional fields  see \cite{CVETICYOUM}).
  Without loss
of generality,  the non-zero charges can be   chosen to be
 ${ P}^{(1)}_1,
{ P}^{(2)}_1,{ Q}^{(1)}_2, {  Q}^{(2)}_2$, {  i.e.},
to be  the  charges of the gauge fields
associated with the first two circles of the six-torus.\foot{
%In
% the notation we  shall be  using
The upper index  refers
to the origin  (Kaluza-Klein metric or two-form)
of the corresponding  $U(1)$ gauge field
and the lower index indicates   the  number of the  internal
circular dimension.}

The generating solution depends only on the four `screened'
charges $({\bf  P}^{(1)}_{1}$,  ${\bf P}^{(2)}_{1},$
${\bf Q}^{(1)}_{2},$   ${\bf Q}^{(2)}_{2})$   $\equiv (
\eta_P P_1^{(1)},$
$\eta_P P_1^{(2)}$,  $ \eta_Q Q_2^{(1)}$,   $
\eta_Q Q_2^{(2)}).$
Here $\eta_{P,Q}=\pm 1$  are
parameters in the Killing spinor constraints which   are chosen
so that $\ \eta_P$ $ {\rm sign}(P_1^{(1)}+P_1^{(2)})= 1\ $ and
$\ \eta_Q$ $ {\rm sign}(Q_2^{(1)}+Q_2^{(2)}) =1$,\  thus yielding a
non-negative
 ADM mass $M_{BPS} = {\bf P}^{(1)}_{1}+{\bf P}^{(2)}_{1} $ $
+{\bf Q}^{(1)}_{2}+{\bf Q}^{(2)}_{2}$  for the four-dimensional
configuration.

The four-dimensional  space-time structure
depends on the  relative signs of the
two magnetic and the two electric
charges.
When the relative signs of the two magnetic
and  the two electric  charges are
the same, we refer to  such  solutions as {\it regular}. They   have  a
 Reissner-Nordstr\" om-type horizon,  null singularity,  or
naked  singularity when  all  four, only three (or two),
or only  one of the charges are nonzero, respectively.
On the other hand,  when the relative
signs for the two magnetic  (and/or two electric) charges are opposite,  we
refer to such  solutions as {\it singular}; they always  have  a naked
singularity
 and repel massive particles \cite{KALLIND,CYHETS}.
In addition, the latter set of configurations become massless
\cite{BEHRNDT,KALLIND,CYHETS}
when the two  screened electric (and two magnetic) charges have opposite
signs and equal magnitudes,  and may contribute to the enhancement of local
 gauge  symmetry \cite{HTII,CYHETS} as well as
local supersymmetry \cite{CYHETS} at special points of moduli space.

The  solution of \cite{CVETICYOUM} was expressed in terms of the fields
appearing in
 the  four-dimensional, low-energy effective action.
In terms of
 the corresponding
 % fundamental
fields of the  ten-dimensional
%$D=10$
 heterotic
string theory (see,  e.g.,  \cite{SEN2})
% for notation)
  $G_{\m\n},\  B_{\m\n},$ $   \Phi, $ $   A^I_\m$  ($\m,\n=0,\cdots,
9$, \ $I=1,\cdots, 16$), { i.e.}, the string-frame
metric, the  two-form field, the
dilaton and the Abelian (Cartan subalgebra) Yang-Mills fields,
the solution takes the form:
%one  finds that
% the non-trivial part of the solution is effectively six-dimensional
%$(t,r, \theta,\vp; y_1,y_2)$,
% with the remaining four toroidal coordinates
%$y_3, ..., y_6$ decoupled,
\be
ds^2_{10}
= \sum_{n=3}^{6} dy_n^2  +  F(r)\left[ 2dtdy_2 +K(r)dy_2^2\right]
\label{sol}
\ee
$$
+ \  f(r)\left\{ k(r)\left[ dy_1 +{\bf
P}^{(1)}_1 (1-\cos\theta )d\varphi\right] ^2
+  { {k\inv (r)}}
\left[ dr^2+r^2 (d\theta^2 +  \sin^2\theta d\varphi^2)\right] \right\}
,  $$
\be
 B_{\vp,1}={\bf P}^{(2)}_1(1-\cos\theta ) , \ \ \
  B_{t, 2}=F(r) , \ \ \  \Phi =
 \ha \ln \big[ F(r) f(r)\big]  ,\  \  \ A^I_\m=0  ,
\label{soll}
\ee
where
\be
F\inv  =  1+{{{\bf Q}^{(2)}_2}\over r}\  ,\ \ \
  K=1+{{{\bf Q}^{(1)}_2}\over r}\ ,  \ \ \
f = 1+{{{\bf P}^{(2)}_1}\over r}\ ,\ \ \
k\inv = 1+{{{\bf P}^{(1)}_1}\over r}\ .
\lab{FKfk}
\ee
Here $t$  and  $(r,\theta,\vp)$ are time  and the spherical coordinates of the
four-dimensional
 % 4-d
 space-time, while  $(y_1,\cdots, y_6)$ are the periodic  coordinates
 of the six-torus.   The radial coordinate $r$  is chosen
in such a way  that the horizon (or singularity)
of the   four-dimensional
%$D=4$
 Einstein-frame metric
for the  regular solutions is
at $r=0$, while the singular solutions have a naked singularity at $r=$
max[min $(|{\bf  P}^{(1)}_{1}|$,  $|{\bf P}^{(2)}_{1}|$),  min $(|{\bf
Q}^{(1)}_{2}|,$   $|{\bf Q}^{(2)}_{2}|)$].
Note that the non-trivial  $(t,r, \theta,\vp; y_1,y_2)$-part of the solution is
six-dimensional.
% with the remaining four toroidal coordinates
%$(y_3, ..., y_6)$ decoupled.
These solutions admit a straightforward multi-center generalization
 (see also below).

It turns out \cite{CVETICYOUM} that the  non-zero magnetic
and non-zero electric charges each break  $1\over 2$
%half
 of the maximal
number of supersymmetries.  Thus, purely electric (or purely magnetic)
configurations preserve
% one half
$1\over 2$,
  while dyonic solutions
preserve only
%one quarter
$1\over 4$ of the  $N=4$ supersymmetry in four dimensions.\foot{The first and
the second sets of configurations  belong to
the vector- and hyper- supermultiplets with highest spins 1 and $3\over 2$
\cite{STRATH,KALL3},
respectively.}
%The important consequence of this is that while
Consequently,  the
 corresponding  full (heterotic) string \sm action
will have only  $N=2$
 world-sheet  supersymmetry.
However,  its transverse  `magnetic'
 part  will have  an increased,
$N=4$,  world-sheet supersymmetry.

\vskip 0.2cm

3.{\it \ Conformal Chiral Null Models with  $N=4$ Supersymmetric Transverse
Part.}\  Our goal is  to write down the world-sheet supersymmetric
heterotic or   type II   superstring  \sm action
corresponding to the background (1)--(3).
The world-sheet action will turn out to correspond to
 a special case of a  generalized chiral null model
which  is  conformally invariant to all orders in $\a'$,
in spite of the fact that the
full \sm action will have  only $N=2$
world-sheet supersymmetry.
%It  will be  crucial for this  that the electric and magnetic charges
%correspond to  two {\it different}
%internal dimensions.
A novel mechanism  of preserving
conformal invariance
 which  will be  at work here is based
on  combining  the chiral
 null structure of the `electric' part of the model
with the extended $N=4$  world-sheet supersymmetry of  its  four-dimensional
transverse `magnetic'  part.
Let us first describe a general class of  such conformal
% In the following we describe that class of exact
chiral null models.

The chiral null models   \ci{TH,TSH,TS}  is a  class of  2d
$\s$-models  which generalize both plane wave type  and fundamental string
type
models
%and have one conserved chiral null current
 \be
   L =  F(x)  \del u \left[\bd v +
   K(u,x) \bd u  +   {\cal A}_i(u,x)  \bd  x^i \right]
 +   (G_{ij} + B_{ij})(x)\ \del x^i \bd x^j    +    {\cal R}
\P (u,x)\  .
 \la{lag}
\ee
Here  $u,v$ are `light-cone' coordinates, $x^i$  are
`transverse space' coordinates
and ${\cal R}\equiv \four \a'
\sqrt{ g^{(2)}}  R^{(2)}$.
% ${\cal R}$ is proportional to the  curvature of the world-sheet metric.
The affine symmetry
$v'=v + h(z)$ implies the existence of a conserved
chiral null current. The corresponding metric has  a null Killing vector
and the  generalized connection  with torsion
has a special holonomy.

 A remarkable property of
 the
model (\ref{lag}) is  that there exists a renormalisation  scheme
in which it is conformal to all
orders in $\a'$ provided\   \
 (i) the `transverse'  \sm  $(G_{ij} + B_{ij})\del
x^i \bd x^j$
is conformal
 when supplemented with a  dilaton coupling
  $\p(x)$  (so that $(G_{ij}, B_{ij},\p) $
represent an exact string solution),\
  and \  (ii) the functions $F\inv,K,{\cal A}_i, \P$ satisfy\foot{In what
follows we shall consider  for simplicity  the special case when
$K, {\cal A}_i$,$\P$ are $u$-independent.
Then the action is also invariant under $v'=v-
\eta(x) , \ {\cal A} '_i= {\cal A}_i  + \del_i
\eta$.}
$$
F^2(-\o F\inv +  \del^i \p \del_i F\inv)=0\ , \ \ \
F( - \o K    +  \del^i \p \del_i K)
=0 \ , \ \ \  \o = \ha  \nabla^2   +  O(\a') \ ,
$$
\be
F( -\ha  \hat \nabla_i {\cal F}^{ij} + \del_i \p
{\cal F}^{ij})  =0  \ ,  \ \     {\cal F}_{ij} \equiv  \del_i {\cal A}_j -
\del_j {\cal A}_i\ ,
\  \  \  \ \ \
 \P =  \p  + \ha \ln F\   .
\la{cond}
\ee
Here    $\o$ is the scalar anomalous dimension operator. In general,  it
contains $(G_{ij},B_{ij})$-dependent corrections to all
orders in $\a'$ (see \cite{calg,TS}).

 The simplest example of the chiral null model
 is the one   with the {\it flat}  transverse space
  $G_{ij}+B_{ij}=\delta_{ij}$ and
 constant (or linear)  dilaton
$\p_0$. At the points where $F$ is
non-vanishing  the conditions (\ref{cond})  then  reduce to
\be
   \del^i\del_i  F\inv=0 \ , \ \ \  \del^i\del_i  K =0 \ , \ \ \   \del_i {\cal
F}^{ij} = 0 \ , \ \ \ \  \P=  \p_0 + \ha  \ln F \ .
\la{lin}
\ee
This model
  describes  a large  class  of
  exact string solutions, in particular, higher dimensional  plane waves,
 fundamental strings and, upon dimension reduction,  four-dimensional
 supersymmetric  electric
black hole-type solutions  \ci{TH,TS,BEHRNDT2,BEHRNDT,HOS}.
%things above are repetition of things in the introduction--this is ok--for
%%%logic of exposition.
The  {\it electric} backgrounds  are embedded into  a higher dimensional model
by assuming that  a linear combination of $u$ and $v$ is a compact
 Kaluza-Klein dimension $y$, { i.e.}, that
the electric field originates from the  `light-cone' ($u,v$) part of the model,
while the transverse spatial part remains  flat.
%At the same time,
On the other hand, in order  to embed the {\it magnetic}
 field,   one has to consider
 models with  non-trivial transverse spatial part.
This suggests that the  {\it dyonic} solutions discussed  above
should be described by  a more  general chiral null
model (\ref{lag}) with a {\it curved } transverse part.

It is indeed
 possible  to  obtain   exact  solutions
 with curved transverse space  when the  conformal field theory
corresponding to the transverse part of (\ref{lag})  is  known  explicitly
\cite{TSH,TS}.\foot{In that case  the  `tachyonic' operator
$\o$ is  given  by the  zero-mode part of the
CFT Hamiltonian  $H $.
Fixing a particular scheme (e.g.,  the one
where $H $ has the standard
`Klein-Gordon with  a  dilaton'  form),
 one is able  to  determine  the  exact form
of  the background fields $(G_{ij},B_{ij},\p)$
{\it and} $F,K,{\cal A}_i$.  This  produces, in particular,
 the exact solutions which are
 `hybrids' of  the gauged Wess-Zumino-Witten (WZW)  and plane wave
  (or fundamental string)  solutions
\cite{TSH}.}
Here we shall consider
 another  important
exactly solvable case,
 when the  transverse  \sm   has {\it extended
$N=4$ }
world-sheet supersymmetry.
 Then one can
 argue, following \cite{NNN,HP,HPP},
that there exists a scheme in which
 not only the transverse \sm couplings
 $(G_{ij},B_{ij},\p$), but also the `tachyonic' operator $\o$
 and thus the functions $F,K,{\cal A}_i,\P$
retain their leading-order form.
In what follows we shall set ${\cal A}_i$-couplings in (\ref{lag})
equal to zero.
As a result,  (\ref{lin}) is replaced by
\be
   \del_i (e^{-2\p}\sqrt G G^{ij} \del_j F\inv) =0  ,  \  \
  \del_i (e^{-2\p}\sqrt G G^{ij} \del_j K) =0  , \ \   \P=  \p(x) + \ha  \ln
F(x)  .
\la{linn}
\ee
As explained in \ci{HPP},
a $(4,0)$ supersymmetric \sm is exactly
conformal  in a special  scheme where  the $(4,0)$ supersymmetry is preserved.
It is
thus sufficient to demand  the   $(4,0)$ supersymmetry of the transverse \sm
 in order to ensure  that the transverse
%Thus each  transverse $(4,0)$ supersymmetric
\sm  gives rise
 to an exact  (`left-right asymmetric')
solution of the heterotic string theory.
The same background will be represented by  $(4,1)$ supersymmetric \sm
in the context of type II superstring
theory (in some special cases
the $(4,1)$ \sm  may have $(4,4)$ supersymmetry \ci{chs}).
The corresponding exact type II  superstring solution
can be reinterpreted also as  another, `symmetric',
heterotic string solution  which is obtained by
embedding the generalized
Lorentz connection   into the gauge group (see, e.g.,
\cite{chs,kalor}).\foot{The resulting
gauge field background is of the next to the leading order in $\a'$
(and  therefore  can be  ignored in  solving the effective field equations),
i.e., the `asymmetric' and `symmetric' heterotic string solutions agree
to the leading order in $\a'$.
The schemes in which these  two  exact heterotic string solutions
are exact
 are related by local covariant
  field redefinitions involving terms of all orders in $\a'$  \ci{HPP,TH}.
}

The   conditions for a heterotic (type II) \sm  to  admit a  $(4,0)$
 ($(4,1)$)
supersymmetry were discussed  in \ci{NNN,NNNN,HP,chs,NNNNN}.
Below we  shall  consider  a particular
 model with  a {\it four}-dimensional
transverse part such that its metric is
\be
G_{ij} = f (x) g_{ij} \ , \  \ \ \ \ \ \ \ \na^2 f =0 \ ,
\la{hye}
\ee
where $g$ is a hyper-K\"ahler metric and $\na\equiv \na (g)$
 \ci{chs}.\footnote{Examples of other  four-dimensional $(4,0)$
models with torsion which have   metrics   not
conformal to  hyper-K\"ahler ones
were considered in \ci{NNNNN}.}
 Then the  torsion and dilaton should   satisfy
$H^{ijk}$  $ =-  {G}^{-1/2}  $ $ \epsilon^{ijkn} \del_n \ln f$ and $\p= \ha \ln
f$.
We shall assume that the four-dimensional
 hyper-K\"ahler metric $g_{ij}$
 has at least one   translational
isometry (in  $x_4$-direction).
Then  this metric can be put into the form ($x_i =(x_s,x_4), \
s=1,2,3 $)
\be
g_{ij} dx^idx^j=  k(x)  \left[dx_4 + a_s (x) d x^s\right]^2
    +  k^{-1} (x) d x_s dx^s\  ,
\la{hype}
\ee
where $k$ and $a_s$ depend on $x_s$ only,  and satisfy the conditions
\be
   \del_s\del^s k\inv =0 \  ,  \ \ \ \ \
\del_{p} a_q - \del_q a_{p}
 = \epsilon_{pqs} \del^s   k\inv\  ,
\la{hypp}
\ee
 which imply  the self-duality of  the curvature of $g_{ij}$
(here $\epsilon_{pqs}$ is the flat-space antisymmetric tensor
and the indices are contracted using $\delta_{pq}$).
Depending on  the asymptotic behaviour   of
$k\inv$
these are (multi-center) Eguchi-Hanson or Taub-NUT
gravitational instantons (see,  e.g., \ci{TNUT}).
Note that  for $g_{ij}$ in  (\ref{hype})
 the condition on $f(x)$ in  (\ref{hye})
becomes simply the `flat' one,  \   $\del_i \del^i f  =0$.\foot{While
$k$ and $a_s$ are assumed to depend  on  $x_s$ only, $f$ may, in general,
 depend also on  $x_4$, thus allowing for a more general set of solutions (see
 below).}

We shall thus consider the
 $(1,0)$ (or ($1,1$))  supersymmetric\foot{The full   heterotic (type II)
model  is
  actually $(2,0)$ ($(2,1)$) supersymmetric \ci{TH},
 in agreement with the $N=1$ four-dimensional space-time supersymmetry
of the corresponding background.}
six-dimensional
 chiral null  model with $(4,0)$ (or ($4,1$)) supersymmetric
\foot{When $g_{ij}$ is flat ($k=1$)  the corresponding  $(4,1)$
type II (or `symmetric' heterotic) `transverse' \sm  is actually
$(4,4)$ supersymmetric \ci{chs}. The same  may be  true
also in the general case of (9),(10).} `transverse' part
   which has the  following
bosonic term in  the Lagrangian
\be
 L = F(x)  \del u \left[\bd v +  K(x) \bd u \right]  + \ha {\cal R} \ln F(x)
+ L_{\bot}\ ,
\la{lagr}
\ee
$$
L_{\bot}=  f(x)\left[ k(x)  \big( \del x_4 + a_s (x) \del x^s\big) \big(
\bd x_4  + a_s (x) \bd x^s\big)
    +  k^{-1} (x) \del x_s \bd x^s \right]
$$
\be
 +\   b_s (x) (\del x_4 \bd x^s - \bd x_4 \del x^s)
  +   {\cal R}   \p(x) \  ,
 \la{latra}
\ee
where, in addition to (\ref{hypp}),  we shall assume that $f=f(x_s)$ and
\be
   \del_s\del^s f =0 \  ,  \ \ \ \
\del_{p} b_q - \del_q b_{p}
 = \epsilon_{pqs} \del^s   f \  ,  \ \ \ \ \ \ \p= \ha \ln f \  .
\la{hpp}
\ee
Choosing  $F$ and $K$  to be  independent of $x_4$  and
observing that  the transverse metric  $G_\bot \equiv$ $ (G_{ij})$ and  the
dilaton in
 (\ref{latra}),(\ref{hpp}) are such that (for {\it  any} $f$ and $k$)
$\ e^{-2\p} \sqrt {G_\bot} G^{pq}_\bot= \delta^{pq} \ (p,q=1,2,3)$,
   we   conclude that the conditions (\ref{linn})
on $F$ and $K$ take  the  {\it flat  space}  form (cf. (\ref{lin}))
\be
\del_s\del^s  F\inv =0 \  ,  \ \ \ \ \ \ \  \ \del_s\del^s  K =0 \ .
\la{liin}
\ee
Their solutions are  thus given by harmonic functions of $x_s$ which
are
{\it independent} of  a  particular choice of the
functions  $f,k,a_s,b_s$ in $L_{\bot}$.
The above conditions  (\ref{hypp}),(\ref{hpp}),(\ref{liin}) are just  the
leading-order (one-loop)  conformal invariance conditions
for the    bosonic \sm (11). However,  there exists a scheme
in which they are {\it exact}  (all-loop)
conditions for the conformal invariance
 of the corresponding $(1,0)$ (or $(1,1)$)
supersymmetric $\s$-model.

To summarize, the $(1,1)$ supersymmetric  model (11),(12)
represents an exact solution of the  type II superstring theory.
There are also two  heterotic string solutions associated
with the \sm \  (11),(12). The `asymmetric' one is obtained
by considering its  $(1,0)$ supersymmetric extension
and arguing that the  chiral null structure combined with the $(4,0)$
supersymmetry of the `transverse' part is sufficient for its
exact conformal invariance in a certain scheme.\foot{ A subtlety that appears
here is  related to the presence of the \sm anomaly
implying the modification $H\to \hat H$ of the torsion $H=dB$
by the Lorentz Chern-Simons term in the expressions for the
conformal anomaly $\b$-functions. This modification
is trivial  in the purely `electric' case $f=k=1$ \ci{TH}
 (where $d\hat H= 2\a' {\rm tr}(R\wedge R)=0$),
  but   is present in the
 `magnetic' case \ci{chs,kalor}.
 Then one is to assume that there exists a deformation
 of $B_{ij}$ by higher order $\a'$-terms
 which solves the conformal invariance equations.
 Because of the underlying $(4,0)$ supersymmetry
 of the transverse model
 it should be possible to `undo' this deformation
 by a local change of a scheme \ci{HP,HPP}.}
The `symmetric' heterotic string solution is obtained
by embedding the {\it `transverse'} (`magnetic')
part of the Lorentz connection $\omega_-$
into the gauge group, i.e., by adding the gauge field background
to make  the  transverse  \sm \
anomaly-free and $(4,1)$ supersymmetric as in the
special cases considered in \ci{chs,kalor}.
Note that  the {\it full} six-dimensional heterotic
 \sm (11),(12) remains only  $(2,0)$ supersymmetric
since  the  embedding into the gauge group is not possible
 for the  `electric' part of the Lorentz connection which
has a  non-compact holonomy \ci{TH}
(such an embedding  is not actually needed for the
exact conformal invariance
\ci{TH}).

If the  harmonic functions $F,K,f,k$
are chosen  so  that they approach constant values at large $x$
(as in the case of (3)), i.e.,  that  at large $x$  the six-dimensional
\sm (11),(12) becomes
a free theory with constant dilaton, then the central charge $c$
of this conformal \sm  has the free-theory value ($c$  is constant
and hence can be evaluated at large $x$).

Note also that the model  (11),(12)
is covariant  under the separate \sm  duality (or target space, $T$-duality)
 transformations
in the two isometric coordinates $u \to \tilde u$  \cite{TH} and
$x_4 \to \tilde x_4$  combined with\foot{We are assuming that $a_s$ and $b_s$
are parallel as in the magnetic case discussed below.
In general, the duality transformation
in $x_4$ produces an extra torsion term $(b_p a_q - a_p b_q) \del x^p \bd x^q$
in the action.}
\be
F \to K\inv\ , \ \   \   K \to F\inv\ ,
\ \ \  \ \ \
f \to k\inv\ , \ \  \   k \to f\inv\ ,
\ \ \  a_s \to b_s\ , \ \  \ b_s \to a_s\  .
\la{dua}
\ee
In the case of the simple harmonic functions in (3) this  transformation is
equivalent to
 ${\bf Q}^{(1)}_2\leftrightarrow {\bf Q}^{(2)}_2$ and
${\bf P}^{(1)}_1\leftrightarrow {\bf P}^{(2)}_1$.

\vskip 0.2cm

4.{\it \  Dyonic BPS  Backgrounds
as Exact Superstring Solutions.} \  Finally, we are
 able to relate the above  class of chiral null models  to the
general class of four-dimensional dyonic BPS saturated states
 described by the
 backgrounds (1)--(3).
The non-trivial
six-dimensional
  part of these backgrounds
 %  superstring action corresponding to (1)--(3)
corresponds to a special case  of the  \sm (\ref{lagr}),(\ref{latra})
where $F\inv , K, f, k\inv $ are four
spherically symmetric one-center
solutions  (3) of the three-dimensional Laplace equation,
and
$$
u=y_2, \ \ v=2t, \ \ x_4=y_1, \  \ dx_s^2= dr^2+r^2 (d\theta^2 +  \sin^2\theta
d\varphi^2),
\ \  r^2=x_s x^s,
$$
\be
 a_s dx^s = {\bf P}_1^{(1)}(1-\cos \t) d\vp\ , \ \  \ \
b_s dx^s = {\bf P}_1^{(2)}(1-\cos \t) d\vp\  .
\la{dee}
\ee
%where the  summation over $s=1,2,3$-indices is  implied.
 Thus,  the model
(11)--(13),(10)
promotes the backgrounds (1)--(3) to {\it exact superstring solutions}.

One   advantage of  having identified
the conformal \sm  behind  the leading-order solution
(1)--(3) is that now  various generalizations
become obvious.
For example, the  multi-center, or `rotating',
%(singular on circles rather than points)
  or `Taub-NUT'-type backgrounds are obtained  by
choosing more general expressions for
the harmonic functions
$F\inv ,$  $ K,$ $  f, k\inv $.
The model  (\ref{lagr}),(\ref{latra}) admits  also
an important extension   which is
found by  assuming that  $f$ may depend also on $x_4$
and  by replacing  the $b_s$-terms   by
$B_{ij} (x) \del x^i \bd x^j$ with the generalized torsion
$\hat H_{ijk}$  $ =- 2\sqrt G G^{nl} \epsilon_{ijkn} \del_l \p $.
Another  possibility  would be to use
a generic $N=4$ supersymmetric
model  as the  transverse part,
 relaxing
the condition of isometry in $x_4$ direction on the
hyper-K\"ahler metric $g_{ij}$ in (\ref{hye}).
These
generalizations yield  new types of  genuine six-dimensional solutions
which are worth further investigation.
The special case  of $F=K=k=1$  and $O(4)$-symmetric $f$
corresponds to the five-brane solution
\cite{chs}.
 One can also consider solutions periodic in $x_4$
(analogs of periodic instantons \cite{HMON})
which  at  distances  large compared to  the period of $x_4$ correspond to
 four-dimensional  backgrounds.

%%%%%%%%%%%%%%%%%%%%%%%%
 For regular solutions with both
  ${\bf Q} _2^{(2)}$ and ${\bf P} _1^{(1)}$  non-vanishing
we can consider the `throat limit'  $r \to 0 $  of the  conformal
model (11),(12),(16),(3).\foot{In this case the dilaton $\Phi=\ha \ln (Ff)$
approaches
a constant for $r\to 0$. Other
cases are related to this one by $T$-duality (15).}  It then
 takes the following simple form
\be
 L_{r \to 0}  =   \left(p \del z \bd z +  q \del \td y_2  \bd \td y_2
 +  2  e^{-z }   \del \td y_2 \bd t  \right)
\ee
$$ +   \   p \left[ \del \td y_1 \bd \td y_1     + \del \vp \bd \vp +
\del \theta \bd \theta  - 2 \cos \theta  \del  \td y_1 \bd \vp \right]  \ ,
$$
where we have introduced the notation:
$z= - \ln r, \ z\to \infty$
and $  \td y_1 = y_1/ {\bf P} _1^{(1)} + \vp , \
 \td y_2 = y_2/{\bf Q} _2^{(2)}, $  \
$p= {\bf P} _1^{(1)} {\bf P} _1^{(2)}, $  \
$q={\bf Q} _2^{(1)} {\bf Q} _2^{(2)}.$
It is easy  to show that
(up to  the  issue  of
periodicities   of coordinates)
the terms in the first bracket
are equivalent to the Lagrangian of the $SL(2,R)$  WZW model
(see, e.g.,  Appendix B of \ci{TH}) while the terms in the second
bracket represent
 the  Lagrangian of the $SU(2)$ WZW model (see, e.g.,  \ci{THR,nels}).
Thus the model  becomes equivalent to the direct product
of   the   $SL(2,R)$ and $SU(2)$ WZW  theories
 divided  by discrete subgroups.\foot{See also \ci{LO} for related
observations. Similar `throat-limit'
 conformal models were discussed in
\ci{THRO}.}
The levels $k$  of the two  WZW models are  both equal to  $4p$.
The
central charge
of this super-conformal  theory is then
 $c= [3(k+2)/k + 3/2] + [3(k-2)/k + 3/2] = 6+3$, i.e., it
is the same  as that of the  theory corresponding to
a flat background.
% flat super-conformal  theory
%defined the region $r\to \infty$.

%%%%%%%%%%%%%%%%%%%%%%%%%%%%%%%%%%%%%%%
%%%%%%%%%%%%%%%%%%%%%%%%%%%%%%%%%%%%%%%%%%%%

Some special cases of the  solution
(11),(12),(\ref{dee}),(3)  obtained
by choosing   special values  of
 the  parameters   $({\bf  P}^{(1)}_{1}$,  ${\bf
P}^{(2)}_{1},$
${\bf Q}^{(1)}_{2},$   ${\bf Q}^{(2)}_{2})$ are:
\begin{itemize}
%%%%%%%%%%%%%%%%%%%%%%%%%%%%%%%%%%%%%%%%%%%%
\item{\it Zero magnetic charges}: \ \
 For ${\bf P}_1^{(1,2)} =0$ ($f=k=1$)
 the  model (\ref{lagr}) is  the chiral null model with  {\it
flat} transverse
part  discussed in detail in \cite{TH,TS} (see also \cite{BEHRNDT,PARK}).
The case  with
${\bf Q}^{(2)}_2=0$ ($F=1$)
  corresponding, from the four-dimensional point
of view,
to the     extreme electric
black hole solution in
Kaluza-Klein theory,
 is  described by
 the  five-dimensional   plane wave  with  flat transverse part
 \cite{GIB}.
The  case  with  ${\bf Q}^{(1)}_2=0$ ($K=1$)
is related to the one with ${\bf Q}^{(2)}_2=0$  by the duality in
$y_2$-direction
(see (\ref{dua})) and describes equivalent four-dimensional
background.
The self-dual case  ${\bf Q}^{(1)}_2= {\bf Q}^{(2)}_2$   ($K= F\inv$)
corresponding
to the extreme electric  $a=1$ dilatonic black hole \cite{GIBMA},  is
represented by   the five-dimensional
fundamental string solution \cite{HRT}.\foot{These
four-dimensional  electric  solutions have the same
type of  charge assignments as certain   elementary string
excitations  and may thus be  identified with them  (see, {  e.g.},
\cite{Duff3}
and references  therein).}
%%%%%%%%%%%%%%%%%%%%%%%%%%%%%%%%%%%
\item {\it Zero electric charges}: \ \
%%%%%%%%%%%%%%%%%%%%%%%%%%%%%%%%%%%%%%%%%%%%%%%
For   ${\bf Q}_1^{(1,2)} =0$ ($F=K=1$)  the  background (1)--(3) is    a
generalized five-brane-type  solution  not studied before.
 The internal space of  the five-brane  (9)
is  a hybrid of the  gravitational
instanton
% (yielding the Kaluza-Klein monopole \cite{GP} in four-dimensions)
and the  $H$-instanton.
%(yielding the $H$-monopole\cite{BANKS,HMON} in
%four-dimensions).
%These solutions in four dimensions
%preserve one  half of the original $N=4$ supersymmetry,  the corresponding
%two-dimensional sigma model possesses at least  $(4,0)$  world-sheet
%supersymmetry and thus corresponds to the exact string
%solution.
The case  with  ${\bf P}^{(1)}_1=0$   ($k=1$) corresponds
to the $O(3)$ symmetric analog of the
five-brane \cite{DL,chs}
whose four-dimensional manifestation is the $H$-monopole \cite{BANKS,HMON,DKL}.
In the   case   with
${\bf P}^{(2)}_1=0$ \  ($f=1$) related to the one with
${\bf P}^{(1)}_1=0$  by  the duality in $y_1$-direction (see (\ref{dua})),
the  space internal to the five-brane describes the
Taub-NUT gravitational instanton (with  the charge
${\bf P}^{(1)}_1$  being quantized in terms of  the radius of $y_2$
to ensure the regularity of the solution).
Its four-dimensional
`image' is the Kaluza-Klein monopole \cite{GP,dukh}.
The  self-dual case
${\bf P}^{(1)}_1= {\bf P}^{(2)}_1$
\ ($f=k\inv$)
corresponds \cite{nels,kalor} to the extreme magnetic $a=1$
dilatonic   black hole in four dimensions
\cite{GIBMA}.
%%%%%%%%%%%%%%%%%%%%%%%%%%%%%%%%%%%%%%%%%%%%%%%
\item {\it Dyonic case}:\ \
The choice   ${\bf Q}^{(2)}_2=0, \ {\bf P}^{(2)}_1=0$
($F=f=1$)
corresponds to  the
 general supersymmetric
Kaluza-Klein black hole solution \cite{CYKK},  which  is described
by the
six-dimensional Brinkmann pp-wave background
 with  curved  transverse space \cite{GIBBP}.
\end{itemize}
%%%%%%%%%%%%%%%%%%%%%%%%%%%%%%%%
%%%%%%%%%%%%%%%%%%%%%%%%%%%%%%%%%%%%%%%

%%%%%%%%%%%%%%%%%%%%%%%%%%%%%%%%%%%%%%%%

5.{\it \  Duality Transformations.} \  We would like to end with a discussion
of
$S$-duality  at the level
of the  six-dimensional conformal  $\s$-model
defined  by (11)--(13),(10),(3)
and  its relation to string-string duality.
Since the electro-magnetic $S$-duality is
 a non-perturbative, four-dimensional
 symmetry,  which
maps (see, e.g., \ci{SEN2}) one  solution of the leading-order
 four-dimensional
low-energy effective heterotic string equations into  the dual one,
  the
 string $\s$-models  corresponding to   the two $S$-dual  solutions
 are  not expected
  to be related in any simple way.
The  above conformal $\s$-model  is  a
 counter-example to this expectation.
 Applying the  $S$-duality transformation
to  the four-dimensional  dyonic solution
(1)--(3)
 one
 gets the  $S$-dual dyonic
 solution with the two  electric  and two magnetic charges
associated with
the internal dimensions $y_1$  and $y_2$, respectively.
This  $S$-dual solution can then be  elevated back to
 the  effectively  six-dimensional solution  (1)--(3),  with the
related  \sm  being again
(11)--(13),  but now with
{\it  $y_1$ and $y_2$  interchanged}.
The transformation $y_1 \leftrightarrow y_2$  and $
 \ {\bf Q}^{(1,2)}_n\leftrightarrow {\bf P}^{(1,2)}_n$, or, more generally,
$u \leftrightarrow x_4 $  and
\be
F \to f\inv\ , \  \  \  \  K \to k\inv \  ,
\   \ \ \ \ \   f \to F\inv \ , \ \ \    \   k \to K\inv  \   ,
\la{nuu}
\ee
is thus the  counterpart
of the four-dimensional $S$-duality
at the  level of the six-dimensional conformal
$\s$-model.\foot{Note that the four-dimensional dilaton corresponding to
(1)--(3)  which is  defined  by
$e^{-2\p_4}$ = $ e^{-2\P}  ({G_{y_1y_1}}  {G_{y_2y_2}})^{1/2}$ =
 $  (KF\inv k f\inv)^{1/2}$
 is invariant under the $T$-duality (15), but
 indeed changes sign under  the $S$-duality $(F,K)$
  $ \leftrightarrow $ $ (f^{-1},k^{-1})$, i.e.,
$ {\bf Q}^{(1,2)}_n $  $ \leftrightarrow $ $  {\bf P}^{(1,2)}_n$.
The same applies to the six-dimensional dilaton  $\P= \ha \ln (Ff)$
(see also below).}

The  {\it six}-dimensional
model (11)--(13) provides a remarkable illustration  of
 a relationship  between
 heterotic--type II   string--string  duality
in {\it six}  dimensions and  $S$-duality  of
heterotic string  compactified on six-torus ($T^6$) \ci{DUF,WITTENII}.
Indeed, the leading terms  of the effective six-dimensional
%in the superstring effection
action of the heterotic string compactified on $T^4$
and  type II string compactified  on $K3$
%in six dimensions
 are  related by
 % (for zero vector fields)
 the  following  (`string-string duality')
transformation  of the metric  $G_{\m\n}$, the
 two-form field  $B_{\m\n}$ and the dilaton $\Phi$
\ci{WITTENII}:
$ G' = e^{-2\P} G$,  \
$ dB' = e^{-2\P} \ast dB, $
\ $\P'=-\P.$
 Applied to the  background fields of the
  six-dimensional   conformal \sm (11)--(13),
representing  a solution in both  dual theories,
this transformation
 maps   the  \sm into
{\it itself}
with
\be
F \to f\inv\ , \    \  \  K \to K\  ,
\   \ \ \ \ \   f \to F\inv \ , \    \  \  k \to k \   ,
\la{duas}
\ee
and $a_s$ and $b_s$  defined by (10),(13).\foot{When applied to
 more general
 $\s$-models   with $x_4$-dependent  functions mentioned above,
the  six-dimensional string-string duality (19)
relates, in particular, the six-dimensional
 fundamental string solution
 \ci{DA}
 ($K=k=1$, \ $f=1$, $F\inv = 1$ + $a/x^2$) and the six-dimensional image
($F=1$, \ $K=k=1,$  $ f= 1 $ + $a/x^2$) of the five-brane solution
\ci{chs} (see  \ci{HAS}).}

Repeating the sequence of string-string duality  transformation    (19) and
$T$-duality  transformation (15) twice
 yields
 (\ref{nuu}) which is
the four-dimensional  electro-magnetic  \ $S$-duality transformation.
This result  is in
 agreement with the  observation
\ci{WITTENII} that
the toroidal $SL(2)_T$ duality  of  type II
string compactified on $K3\times T^2$
implies  the $SL(2)_S$ duality of heterotic string compactified on $T^6$.

Interestingly, the  self-dual case of the  \sm (11),(12),
corresponding to the fixed  point  $F=K\inv = f\inv=k$
(and $a_s=b_s$) of the duality transformations  (15),(18),(19),
describes  in four dimensions   \ci{CVETICYOUM}
the extreme   Reissner-Nordstr\" om solution. The corresponding  \sm has
a  particularly  simple  structure:\foot{Note  that
this theory has {\it two}
chiral currents or affine symmetries:
not only in  the $y_2=v$  direction
  (as in  the general case of (11)) but also in  the $y_1=x_4$ direction.
 This is due to  the special $T$-self-dual choice of  $f=k\inv, \ a_s=b_s$.}
\be
 L =    \del y_1 \bd y_1  + \del y_2  \bd y_2
+  2 a_s (x) \del y_1 \bd x^s   +  2 F(x)  \del y_2 \bd t
 +  F^{-2} (x)  \del x_s \bd x^s ,
\ee
where we have used  the notation in (16), i.e., $y_1=x_4,\ y_2=u,\ t=\ha v$.
The electro-magnetic  duality  corresponds to
interchanging the third and fourth terms
and $y_1 \leftrightarrow y_2$.
This  model is a special case of
\be
 L =    \del y_n \bd y^n   +  2 A^n_{\mu}(x)  \del y_n \bd x^{\mu}
 +  g_{pq} (x)  \del x^p \bd x^q\  ,
\ee
where  $g_{pq}= F^{-2}\delta_{pq}$ and
$A^1_\mu= (0,a_s)$ is  the
magnetic and $A^2_\mu=(F,0)$ is the electric field
($x^\mu= (t, x^s)).$

%%%%%%%%%%%%%%%%%%%%%%%%%%%%%%%%%%%%%%%%%%%%%%%%%%%%%%%%%%%
In conclusion, we
 would like to emphasize that
the  six-dimensional conformal \sm  (11),(12)
 is  the {\it  minimal}  one  which  enjoys the property of being  covariant
with respect to both
the target space  $T-$duality (15) and the electro-magnetic  $S-$duality
(18) or (19).
Indeed,
one needs  {\it two }   charges associated with each  internal dimension
to have covariance under $T$-duality, i.e.,  scalar duality in two dimensions,
and {\it two }  internal dimensions to embed both electric and magnetic
charges, i.e.,  to have  electro-magnetic or  vector duality in four
dimensions.

%%%%%%%%%%%%%%%%%%%%%%%%%%%%%%%%%%%%%%%%%%%%%%%%%%%%%%%%%%%
%\acknowledgments

{\bf Acknowledgements}

We would like to thank J. Gauntlett, G.
Gibbons, G. Horowitz, C. Hull,
R. Khuri, K. Sfetsos,
 P. Townsend, E. Witten and D. Youm for useful discussions.
The work of M.C.  is supported  by  the Institute for Advanced Study funds and
J. Seward
Johnson foundation,  U.S. Department of Energy Grant No.
DOE-EY-76-02-3071, and the National Science
Foundation Career Advancement Award PHY95-12732.
The work of A.A.T.   is supported  by
 PPARC  and
 ECC grant SC1*-CT92-0789.
We also acknowledge  the support of
the NATO collaborative research grant CGR 940870.

\vskip2.mm

%\begin{references}

%\end{references}
\end{document}